\documentstyle[12pt,epsf]{article}
\renewcommand{\baselinestretch}{1.0}

\newcommand{\be}{\begin{equation}}
\newcommand{\ee}{\end{equation}}
\begin{document}
\topmargin 0pt
\oddsidemargin=-0.4truecm
\evensidemargin=-0.4truecm
\renewcommand{\thefootnote}{\fnsymbol{footnote}}

\newpage
\setcounter{page}{1}
\begin{titlepage}
\vspace*{-2.0cm}
\begin{flushright}
\vspace*{-0.2cm}

\end{flushright}
\vspace*{0.5cm}

\begin{center}
{\Large \bf The MSW effect and Solar Neutrinos  
\footnote{Invited talk  given at  the 11th workshop on Neutrino Telescopes, 
Venice, March 11- 14, 2003.}}
\vspace{0.5cm}

{A. Yu. Smirnov$^{2,3}$\\

\vspace*{0.2cm}
{\em (2) The Abdus Salam International Centre for Theoretical Physics,
I-34100 Trieste, Italy }\\
{\em (3) Institute for Nuclear Research of Russian Academy
of Sciences, Moscow 117312, Russia}

}
\end{center}

\begin{abstract}

The MSW (Mikheyev-Smirnov-Wolfenstein) effect is the effect of transformation of one neutrino
species (flavor) into another one in a medium with varying density. 
Three basic elements of the effect include: the refraction of neutrinos in matter,
the resonance (level crossing)  and the adiabaticity.  
The key notion is   {\it the neutrino eigenstates} in matter. 
Physical picture of the effect  is 
described in terms of the flavors and the relative phases 
of  eigenstates and  the transitions between  eigenstates. 
Features of the large mixing realization of the MSW effect 
are discussed. The  large mixing  MSW effect (LMA)  provides the 
solution of the solar neutrino problem. We show in details how this mechanism works.
Physics beyond the LMA solution  is discussed. 
The lower $Ar$-production rate (in comparison with the  LMA prediction)
and absence of significant turn up of the spectrum at low energies 
can be due to an additional effect of  the light 
sterile neutrino with very small mixing.

\end{abstract}
\end{titlepage}
\renewcommand{\thefootnote}{\arabic{footnote}}
\setcounter{footnote}{0}
\renewcommand{\baselinestretch}{0.9}

\section{Introduction}

\subsection{Context}

The key components of the context in which
the mechanism of resonance flavor conversion has been proposed 
include

\begin{itemize}

\item 

Neutrino mixing and oscillations invented by 
B. Pontecorvo (1957, 1958) (neutrino-antineutrino oscillations,  maximal mixing)~\cite{pontosc} 
and  $\nu_e - \nu_{\mu}$ (flavor) mixing as well as  the ``virtual $\nu_e - \nu_{\mu}$  transmutation" 
proposed by  Z. Maki, M. Nakagawa, S. Sakata~\cite{mns} (1962).

\item

Spectroscopy of solar neutrinos: 
the program  put forward by J. N.  Bahcall~\cite{bah}  and independently by 
G. Zatsepin and V. Kuzmin~\cite{zk} 
to study interior of the Sun by measuring fluxes of all components of the 
solar neutrino spectrum. 
It was proposed to perform  several  experiments with different energy thresholds. 

\item

The results of the Homestake experiment~\cite{hom}: they led to formulation 
of the solar neutrino problem which  has  triggered major experimental 
and theoretical developments in neutrino physics in the last 30 years. 
In fact, the problem was predicted by B. Pontecorvo~\cite{pontprob}, who also 
suggested its vacuum oscillation solution~\cite{pontprob,gp} 
(averaged vacuum oscillations with maximal or near maximal mixing).    

\item

Matter effects on neutrino oscillations  introduced by  
L. Wolfenstein~\cite{w1}.

\end{itemize}

\subsection{References}

Here I give (with some comments) references to the early papers on the MSW effect written by  
W.\cite{w1,w2,w3}  and M.-S.\cite{ms1,ms2,ms3,ms4}.\\

\noindent
L. Wolfenstein:\\
\noindent
[1] ``Neutrino oscillations in matter", Phys. Rev. D17:2369-2374, 1978. 
Topics include:  neutrino refraction, mixing in matter, 
eigenstates for propagation in matter, evolution equation, modification of vacuum oscillations.\\

\noindent
[2] ``Effect of Matter on Neutrino Oscillations",  in   
{\it ``Neutrino -78"}, Purdue Univ. C3 - C6, 1978.
The adiabatic formula has been given for massless neutrino conversion in varying density. \\

\noindent
[3] ``Neutrino oscillations and stellar collapse",  
Phys. Rev. D20:2634-2635, 1979.  
Suppression of oscillations in  matter of the  star is emphasized. \\

We  started to work in the beginning of 1984, when Stas Mikheyev had shown me the 
Wolfenstein's paper [1]. The question was about validity of the results 
and necessity to use them in the oscillation analysis of the atmospheric neutrino data 
from the Baksan telescope. \\

\noindent
S.P. Mikheev, A.Yu. Smirnov:\\ 
\noindent
[4] ``Resonant amplification of neutrino oscillations in matter
and spectroscopy of solar neutrinos", Yad. Fiz. 42:1441-1448, 1985
[Sov.J. Nucl. Phys. 42:913-917, 1985.]\\ 
\noindent
[5]  ``Resonant amplification of neutrino oscillations 
in matter and solar neutrino spectroscopy",  Nuovo Cim. C9:17-26, 1986. 
Appearance of these two papers with very close  but not  identical 
content ({\it e.g.}, in [5] we comment on  the effect in the three neutrino context) 
is a result of problems with publications.\\

\noindent
[6] ``Neutrino oscillations in a variable density medium and neutrino bursts
due to gravitational collapse of stars", 
Zh. Eksp. Teor. Fiz.91:7-13, 1986, [Sov. Phys. JETP 64:4-7,1986.].  
Theory of the adiabatic neutrino conversion  is presented.  
Here formulas for  adiabatic probabilities can be found. 
To ``cheat"  editors and referees  and to  avoid a fate of previous papers 
we have removed the term ``resonance" and ``solar neutrinos" as well as  
references  to our previous papers [4] [5]. 
The paper had been submitted to  JETP Letters  in the fall of 1985 
and successfully ... rejected. 
It has been  resubmitted to JETP in December of 1985. 
The theory  is  applied,  of course, 
to solar neutrinos and the paper was  reprinted in 
``Solar Neutrinos: The first Thirty Years", Ed. J. N. Bahcall, et al., Addison-Wesley 1995. \\

\noindent
[7] ``Neutrino oscillations in Matter with Varying density",  
Proc.  of the 
{\it 6th Moriond Workshop on Massive Neutrinos in 
Astrophysics and Particle Physics} (Tignes, Savoie, France) January 25 - February 1,  1986,  
eds. O. Fackler and J. Tran Thanh Van, p. 355 - 372. 
In  two talks at Moriond,  I have 
summarized all our results (excluding solar neutrinos) obtained in 1985.  
The paper contains  (apart from theory of the adiabatic conversion) calculations 
of the Earth matter effect on the solar and atmospheric neutrinos, the graphic representation of 
oscillations and  adiabatic  conversion, some attempts to apply the matter effects to neutrinos in the 
Early Universe etc.. 

One important contribution both to physics of effect  and to its promotion: 
in  Summary talk of Savonlinna  workshop, where our results 
have been presented for the first time,  N. Cabibbo~\cite{cab} 
has given interpretation of the effect in terms of 
eigenvalues and level crossing phenomenon (complementary to our description in terms of eigenstates).   
A  possibility of such an  interpretation 
was mentioned before by V. Rubakov (private communication) and later has been  
developed independently by H. Bethe~\cite{bet}. 

What was in between  1979 and 1985?  
Several papers has been published on neutrino oscillations in matter with constant 
density \cite{bar,pak,haub,raman,paul}. 
In particular, in the paper by V. Barger et al,~\cite{bar} and S. Pakvasa~\cite{pak}, 
it was shown that matter can enhance oscillations  
and for certain  energy the mixing can become maximal.  Furthermore,  matter 
distinguishes neutrinos and antineutrinos and resolves the ambiguity 
in the sign of $\Delta m^2$. In \cite{paul} the index of refraction of neutrinos has been derived 
for moving and polarized medium, correct sign of the matter potential obtained.

\section{Flavors, masses, mixing and oscillations} 

\subsection{Introducing mixing}

The {\it flavor} neutrino states:  $\nu_f = (\nu_e, \nu_{\mu}, \nu_{\tau})$ 
are defined as the states which correspond  to certain charge leptons:  
$e$, $\mu$ and $\tau$. The correspondence is  established by  
interactions: $\nu_{l}$ and $l$ ($l = e, \mu, \tau$)  interact in pairs, 
forming the charged currents. 
It is not excluded that additional neutrino states, the sterile neutrinos, $\nu_s$, exist. 
The neutrino {\it mass states},  $\nu_1$, $\nu_2$, and $\nu_3$,  with masses $m_1$, $m_2$, $m_3$ 
are the eigenstates of mass matrix as well as the eigenstates of 
the total Hamiltonian in vacuum. 

The {\it vacuum mixing}  means that the flavor states do not coincide with the mass eigenstates.  
The flavor states are  combinations of the mass eigenstates: 
\be 
\nu_{l} = U_{l i} \nu_{i}, ~~~ l = e, \mu, \tau, ~~~i = 1, 2, 3,  
\label{mixing}
\ee
where the mixing parameters  $U_{l i}$ form the  P-MNS mixing matrix. 

In the case of two neutrino mixing $\nu_e - \nu_a$, where $\nu_a$ is the 
non-electron neutrino state,  we can write:   
\begin{eqnarray}
\nu_e = \cos\theta~ \nu_1 + \sin \theta~ \nu_2, ~~
\nu_a = \cos\theta ~\nu_2 - \sin \theta~ \nu_1.
\label{2nu}
\end{eqnarray}
Here $\theta$ is the vacuum mixing angle. 
In the three neutrino context $\nu_e$ mixes with $\nu_a$  
in the mass eigenstates $\nu_1$ and $\nu_2$  
relevant for the solar neutrinos,  and $\nu_a$ is maximal or nearly maximal mixture 
of $\nu_{\mu}$ and $\nu_{\tau}$.

\subsection{Two aspects of mixing. Portrait of electron neutrino}

There are two important physical aspects of  mixing. According to 
(\ref{2nu}) the flavor neutrino states are  combinations of the mass eigenstates. 
One can think in terms of  wave packets.  Propagation of 
$\nu_e$ ($\nu_a$) is described a system of two wave packets which correspond to 
$\nu_1$ and $\nu_2$. 

\begin{figure}[htb]
\hbox to \hsize{\hfil\epsfxsize=15cm\epsfbox{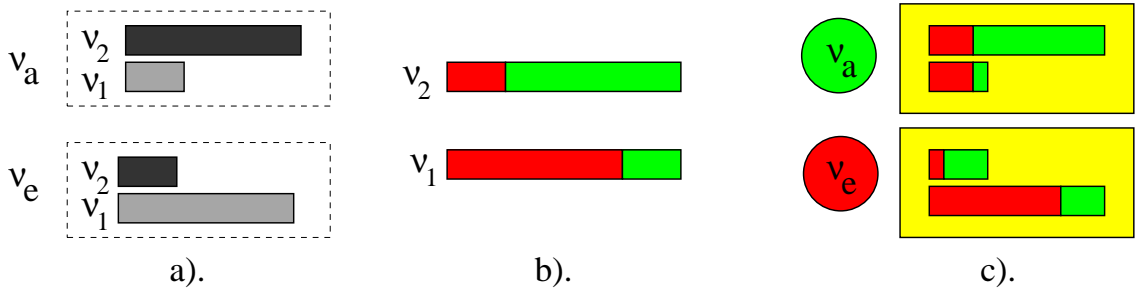}\hfil}
\caption{a). Representation of the flavor neutrino states   
as the combination of the mass eigenstates. 
The length of the box gives the admixture of (or probability to 
find) corresponding mass state in a given flavor state. 
(The sum of the lengths of the boxes is normalized to 1. 
b). Flavor composition of the mass eigenstates. The electron flavor is shown by red 
(dark) and the non-electron flavor by green  (grey). The sizes of the red and green parts give the 
probability to find the electron and non-electron neutrino in a given mass state.
c). Portraits of the electron  and non-electron neutrinos: shown are representations 
of the electron and non-electron neutrino states  as combinations of the eigenstates 
for which, in turn, we show  the flavor composition.
}
\label{mix}
\end{figure}

In fig.~\ref{mix}a). we show representation of $\nu_e$ and $\nu_a$ as the combination of mass states. 
The lengths of the boxes,  $\cos^2\theta$ and $\sin^2\theta$,  give 
the {\it admixtures } of $\nu_1$ and $\nu_2$ in $\nu_e$ and $\nu_a$.  

The key point is that the flavor states are {\it coherent} mixtures (combinations) of the 
mass eigenstates. The {\it relative phase} or phase difference 
of $\nu_1$ and $\nu_2$ in, $\nu_e$ and $\nu_a$ is fixed:   
according to  (\ref{2nu}) it is zero in $\nu_e$ and $\pi$  in $\nu_a$.  Consequently,  there are 
certain {\it interference} effects between $\nu_1$ and $\nu_2$ which depend on the relative 
phase. \\

The relations (\ref{2nu}) can be inverted: 
\be
\nu_{1} = \cos\theta~\nu_e - \sin \theta~\nu_{a}, ~~
\nu_{2} = \cos\theta~\nu_{a} + \sin \theta~\nu_{e}.
\label{2nuin}
\ee
They determine the {\it flavor composition} of the mass states (eigenstates of the Hamiltonian), 
or shortly, the flavors of eigenstates. According to  (\ref{2nuin}) 
a probability  to find the electron flavor in $\nu_{1}$ is given by $\cos^2\theta$, whereas 
the probability  that $\nu_{1}$ appears as $\nu_a$ equals $\sin^2\theta$. This flavor decomposition is 
shown in fig.~\ref{mix}b). by colors. 

Inserting the flavor decomposition of mass states in the representation of the flavors states,  
we get  the  ``portraits" of the electron and non-electron neutrinos fig.~\ref{mix}c). 
According to this figure,  $\nu_e$ is a system of the two mass eigenstates which in turn 
have a composite flavor. 
On the first sight the portrait has a paradoxical feature: there is the 
non-electron (muon and tau) flavor 
in the electron neutrino!  The paradox has a simple resolution: in the $\nu_e$- state  
the $\nu_a$-components of  $\nu_{1}$ and $\nu_{2}$ are equal and have opposite phases. 
Therefore they cancel each other and 
the electron neutrino has pure electron flavor as it should be. 
The key point is  interference: the interference of the non-electron parts is destructive in $\nu_e$. 
The electron neutrino has a ``latent"  non-electron component 
which can not be seen due to particular phase arrangement. However during propagation 
the phase difference changes and the cancellation disappears. This leads to an appearance of the 
non-electron component 
in  propagating neutrino state which was originally produced as the electron neutrino. This 
is the mechanism of neutrino oscillations. Similar consideration holds for the $\nu_a$  state.

\subsection{Neutrino oscillation in vacuum}

In vacuum the neutrino mass states are the eigenstates of the Hamiltonian. 
Therefore dynamics of propagation has the following features: 

\begin{itemize}

\item
Admixtures of the eigenstates (mass states) in a given neutrino state do not change. 
In other words, there is no  $\nu_1 \leftrightarrow \nu_2$ transitions.  
$\nu_1$ and  $\nu_2$ propagate independently.    
The admixtures are determined by mixing in a production point 
(by $\theta$,  if pure  flavor state is produced). 

\item 
Flavors of the eigenstates  do not change. They are also determined by $\theta$. 
Therefore the picture of neutrino state (fig. 1c) does not change during propagation. 

\item 
Relative phase (phase difference) of the eigenstates monotonously increases. 

\end{itemize}
Due to  difference of masses, the states  $\nu_1$ and  $\nu_2$ have different phase 
velocities 
\be
\Delta v_{phase} \approx \frac{\Delta m^2}{ 2E}, ~~~~ \Delta m^2 \equiv m_2^2 - m_1^2, 
\ee
and the phase difference changes as
\be
\Delta \phi = \Delta v_{phase} t .  
\ee
The phase is the only operating degree of freedom here. 

Increase of the phase leads to oscillations. Indeed, the change of  phase  
modifies  the interference: in particular, cancellation of the non-electron   
parts in the state produced as $\nu_e$ disappears and the non-electron component becomes observable.  
The process is periodic: when $\Delta \phi = \pi$,  the interference of non-electron parts is constructive 
and at this point the probability to find $\nu_{\mu}$ is maximal. Later, when $\Delta \phi = 2\pi$, 
the system returns to its original state: $\nu(t) = \nu_e$. 
The oscillation length is the distance at which this return occurs: 
\be
l_{\nu} = \frac{2\pi}{v_{phase}} = \frac{4 \pi E}{\Delta m^2}. 
\ee

The depth of oscillations  is determined by the mixing angle. 
It is given by maximal probability to observe the  ``wrong" flavor $\nu_a$. From the fig. 1c. 
one finds immediately (summing   up the parts with the non-electron flavor in the amplitude) 
\be
depth~~ of~~ oscillations~ = \sin^2 2\theta. 
\ee

The oscillations are the effect of the phase increase which changes the interference pattern. 
The depth of oscillations is the measure of mixing. 
 
\section{Matter effect.}

\subsection{Refraction}

In matter,  neutrino propagation  is affected by interactions. At low 
energies the  {\it  elastic forward scattering} is relevant only 
(inelastic scattering can be neglected) \cite{w1}.   
It can be described  by the potentials $V_e$, $V_a$. In usual medium 
a difference of the  potentials for $\nu_e$ and $\nu_a$ 
is due to  the charged current scattering 
of  $\nu_e$ on electrons  ($\nu_e e \rightarrow \nu_e e$) \cite{w1}: 
\be
V = V_e - V_a  = \sqrt{2} G_F n_e~,
\ee
where $G_F$ is the Fermi coupling constant and $n_e$ is the number density of electrons. 
Equivalently, one can describe the effect of medium in terms of the refraction index: 
\be
n_{ref} - 1 = \frac{V}{p}. 
\ee  
The difference of the potentials leads to an appearance of  
additional phase difference in the neutrino system: 
$\Delta \phi_{matter} \equiv (V_e - V_a) t$. 
The difference of potentials (or refraction indexes) determines 
the {\it refraction length}: 
\be 
l_0 \equiv \frac{2\pi}{V_e - V_a} = \frac{\sqrt{2}\pi} {G_F n_e} . 
\ee
$l_0$  is the distance over   which an additional ``matter" phase  equals $2\pi$. \\

In the  presence of matter the Hamiltonian of system changes: 
\be
H_0  \rightarrow H = H_0 + V ,
\ee
where $H_0$ is the Hamiltonian in vacuum.  Correspondingly, the eigenstates and the eigenvalues 
change: 
\begin{eqnarray}
\nu_1,~~ \nu_2 ~~~\rightarrow ~~~\nu_{1m}, ~~\nu_{2m},\\
\frac{m_1^2}{2E},~~ \frac{m_2^2}{2E}~~ \rightarrow ~~ H_{1m},~~ H_{2m} . 
\end{eqnarray}

The mixing in matter is determined with respect to the eigenstates in matter  $\nu_{1m}$ and $\nu_{2m}$. 
Similarly to (\ref{2nu}) the mixing angle in matter, $\theta_m$,  gives  the relation between  
the eigenstates in matter and  the flavor states: 
\be
\nu_e = \cos\theta_m \nu_{1m} + \sin \theta_m \nu_{2m}, ~~
\nu_a = \cos\theta_m \nu_{2m} - \sin \theta_m \nu_{1m}.
\label{2num}
\ee
Furthermore, in matter both the eigenstates and the eigenvalues, and consequently, 
the mixing angle depend on matter density and neutrino energy. 
It is this dependence activates new degrees of freedom of the system and 
leads to qualitatively new effects. 

\subsection{Resonance. Level crossing}

\begin{figure}[htb]
\hbox to \hsize{\hfil\epsfxsize=10cm\epsfbox{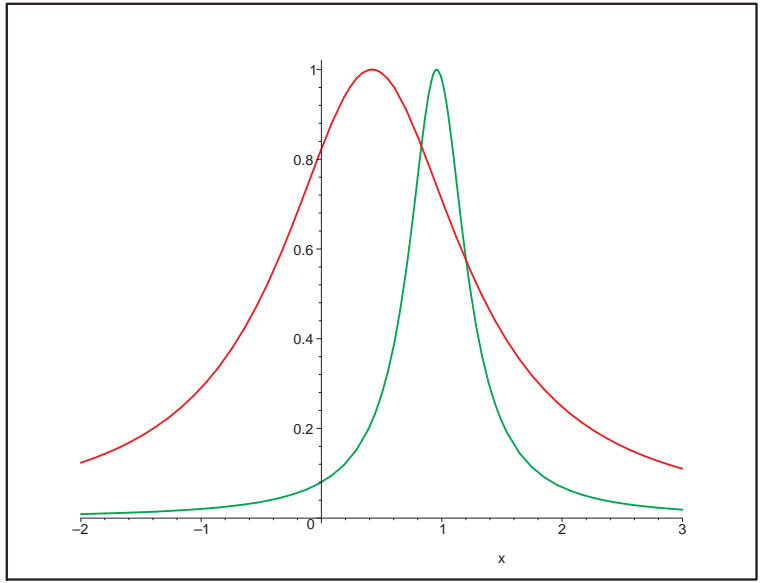}\hfil}
\caption{The dependence of the  effective mixing parameter $\sin^2 2\theta_m$ on 
the ratio $x = l_{\nu}/l_0$ for two different values of the vacuum mixing: $\sin^2 2\theta = 0.825$ 
(red) which  corresponds to the LMA solution and  $\tan^2 \theta = 0.08$ (green) which is 
at the upper bound on 1-3 mixing. The semi-plane  $x < 0$ corresponds to the antineutrino channel. 
}
\label{fres}
\end{figure}

In fig.~\ref{fres} we show  dependence of the effective mixing parameter in matter,  
$\sin^2 2\theta_m$, on ratio of the oscillation and refraction lengths: 
\be
x \equiv \frac{l_{\nu}}{l_0} = \frac{2 E V}{\Delta m^2} \propto E n_e
\ee
for two different values of vacuum mixing angle. 

The dependence in fig.~\ref{fres} has a resonance character. At 
\be
l_{\nu} = l_0 \cos 2\theta~~~~~~~{\rm  (resonance~~ condition)} 
\label{res}
\ee
the mixing becomes maximal: $\sin^2 2\theta_m = 1$. 
For small vacuum mixing the condition (\ref{res}) reads: 
\be
Oscillation~~ length \hskip 0.5cm \approx \hskip 0.5cm Refraction~~ length. 
\ee
That is, the eigen-frequency which characterizes a  system of mixed neutrinos, 
$1/l_{\nu}$,  coincides with the eigen-frequency of medium, $1/l_0$. 
 
For large vacuum mixing (for solar  LMA: $\cos 2\theta = 0.4 - 0.5$) there is a significant 
deviation from the equality. Large vacuum mixing corresponds 
to the case of strongly coupled system  for 
which, as usual, the shift of frequencies occurs.

The resonance condition  (\ref{res}) determines the resonance density: 
\be
n_e^{R} = \frac{\Delta m^2}{2E} \frac{\cos 2\theta}{\sqrt{2} G_F}~.
\label{eq:resonance}
\ee
The width of resonance on the half of the height (in the density scale) is given by 
\be
2 \Delta n_e^R = 2 n_e^R \tan 2\theta, 
\label{width}
\ee
Similarly, one can introduce the resonance energy and the width of  
resonance in the energy scale. 
The width (\ref{width}) can be rewritten as 
\be
\Delta n_e^R = n_0 \sin 2\theta,~~~~~ n_0 \equiv  \frac{\Delta m^2}{2 \sqrt{2} E G_F}~.
\ee
When the vacuum  mixing  approaches maximal value,   
the resonance shifts to  zero density: $n_e^R \rightarrow 0$, the width of the resonance increases 
converging to fixed value: $\Delta n_e^R \rightarrow   n_0$. 

In medium with varying density, the layer where 
the density changes in the interval 
\be
n_e^R \pm \Delta n_e^R   
\ee
is called the resonance layer. 

\begin{figure}[htb]
\hbox to \hsize{\hfil\epsfxsize=11cm\epsfbox{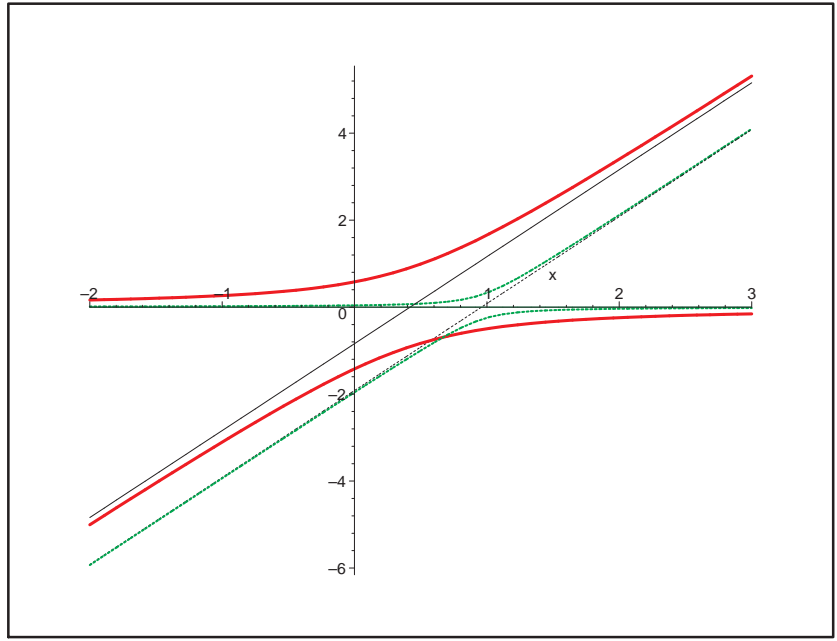}\hfil}
\caption{Level crossing scheme. Dependence of the eigenvalues 
of the Hamiltonian in matter, $H_{1m}$ and $H_{2m}$,  on the ratio  $x \equiv l_{\nu}/l_0$ for 
two different values of  vacuum mixing $\sin^2 2\theta = 0.825$ (solid, blue lines)
and $\sin^2 \theta = 0.08$ (dashed, red lines).
}
\label{fcross}
\end{figure}

In fig.~\ref{fcross}  we show  dependence of the eigenvalues $H_{im}$ on the ratio 
$l_{\nu}/{l_0}$ (level crossing scheme) \cite{cab,bet}. 
In  resonance,  the level splitting is minimal and 
therefore the oscillation length being inversely 
proportional the level spitting,  is maximal. 

The resonance  has physical meaning both for small and for large  mixing.  
Independently of vacuum mixing in the resonance 

\begin{itemize}

\item 
the flavor mixing is maximal;
 
\item
the level splitting is minimal; correspondingly, 
in uniform medium the oscillation length is maximal; 

\end{itemize}

When the density changes on the way of neutrinos,  
it is in the resonance layer the flavor transition mainly occurs.

\subsection{Degrees of freedom. Two effects}

An arbitrary neutrino state can be expressed in terms of the
instantaneous eigenstates of the Hamiltonian, $\nu_{1m}$ and  $\nu_{2m}$,  as
\be
\nu (t) = \cos\theta_a \nu_{1m} + \sin \theta_a \nu_{2m} e^{i\phi}~,
\label{exp}
\ee
where
\begin{itemize}

\item
$\theta_a = \theta_a (t)$ determines the  admixtures of  
eigenstates in $\nu (t)$;\\

\item
$\phi(t)$ is the phase difference between the two eigenstates (phase
of oscillations):
\be
\phi(t) = \int_0^t \Delta H dt' + \phi(t)_T~,
\ee  
here $\Delta H \equiv H_{1m} - H_{2m}$. 
The integral gives the adiabatic phase
and $\phi(t)_T$ is the rest which can be related to
violation of adiabaticity. It may also have a
topological contribution  (Berry phase) in more complicated systems;\\

\item 
$\theta_m(n_e(t))$ determines the  flavor content of the eigenstates:   
$\langle \nu_e| \nu_{1m}\rangle = \cos \theta_m $, {\it etc.}.

\end{itemize}

Different processes are associated with these three 
different degrees of freedom. In what follows we will consider two of them:   

1. The resonance enhancement of neutrino oscillations which 
occurs in matter with constant density. It is induced by 
the relative phase of neutrino eigenstates. 

2. The adiabatic (partially adiabatic) conversion which occurs in medium 
with varying density and is related  to the  change of mixing or flavor of the neutrino 
eigenstates.

In general, an interplay of the oscillations and the resonance conversion occurs. 

\section{The MSW effect}

\subsection{Oscillations in matter. Resonance enhancement of oscillations}

In medium with constant density the mixing is constant: $\theta_m (E, n) = const$. 
Therefore

\begin{itemize}

\item
The flavors of the eigenstates do not change. 

\item
The admixtures of the eigenstates do not change.  
There is no $\nu_{1m} \leftrightarrow \nu_{2m}$ transitions, 
$\nu_{1m}$ and  $\nu_{2m}$ are the eigenstates of propagation. 

\item
Monotonous increase of the phase difference between the eigenstates occurs: 
$\Delta \phi_m = (H_{2m} - H_{1m}) t$. 

\end{itemize}

This is  similar to what happens in vacuum.  
The only operative degree of freedom is  the phase. Therefore,  as in  vacuum, 
the evolution of neutrino  has a character of oscillations. 
However,  parameters of oscillations (length, depth)  differ from the parameters in vacuum. 
They are determined by the mixing in matter and by the effective energy splitting in matter: 
\be
\sin^2 2\theta   \rightarrow     \sin^2 2\theta_m, ~~~~~~  
l_{\nu} \rightarrow  l_{m} = \frac{2\pi}{H_{2m} - H_{1m}}. 
\ee

For a given density of matter the parameters of oscillations depend on the neutrino energy 
which leads to a characteristic  modification of the energy spectra. 
Suppose a source produces the $\nu_e$- flux $F_0(E)$. 
The flux crosses a layer of length, $L$,  
with a constant density $n_e$  
and then detector measures the electron component of the flux 
at the exit from the layer, $F (E)$. 
\begin{figure}[htb]
\hbox to \hsize{\hfil\epsfxsize=7cm\epsfbox{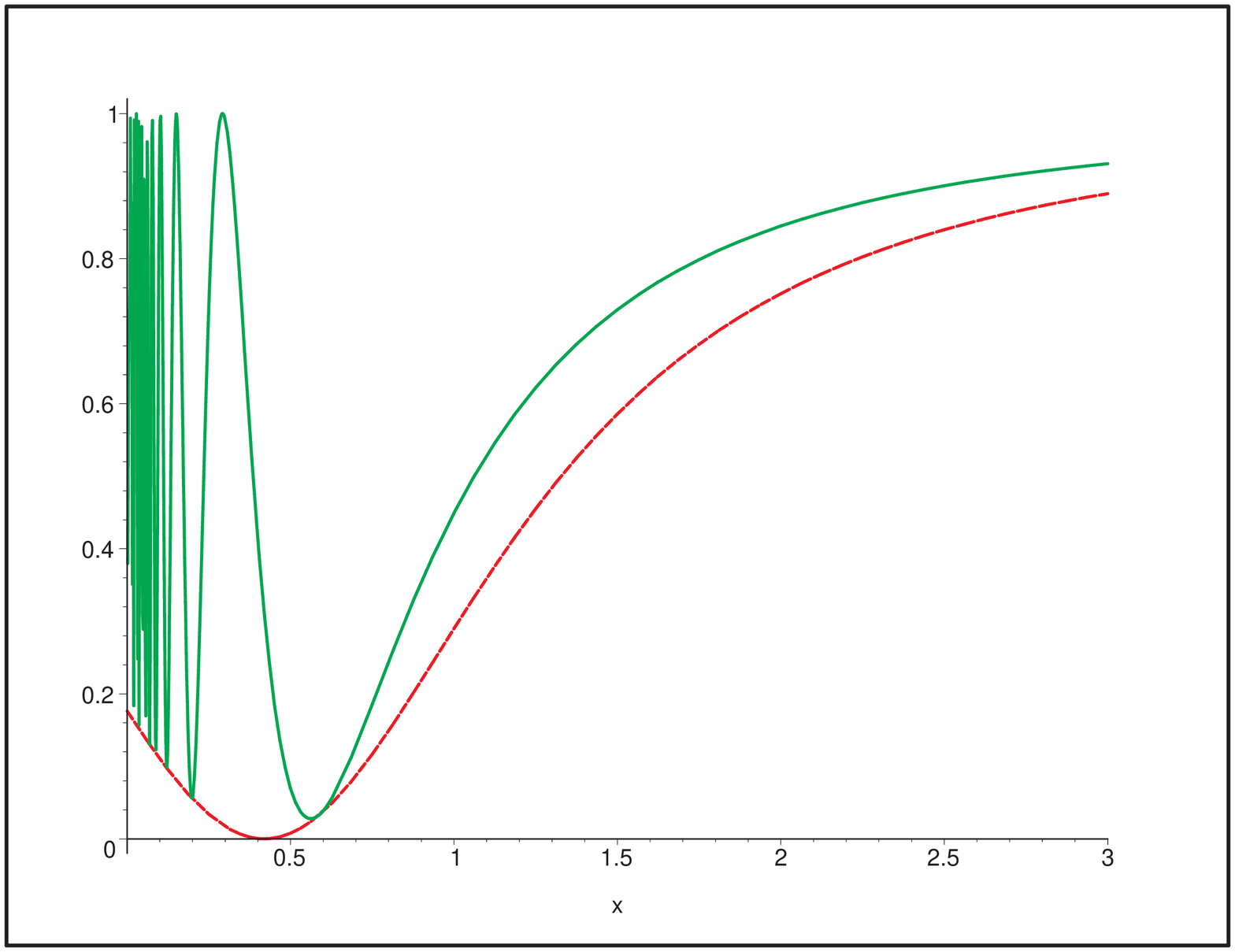}\hfil\epsfxsize=7cm\epsfbox{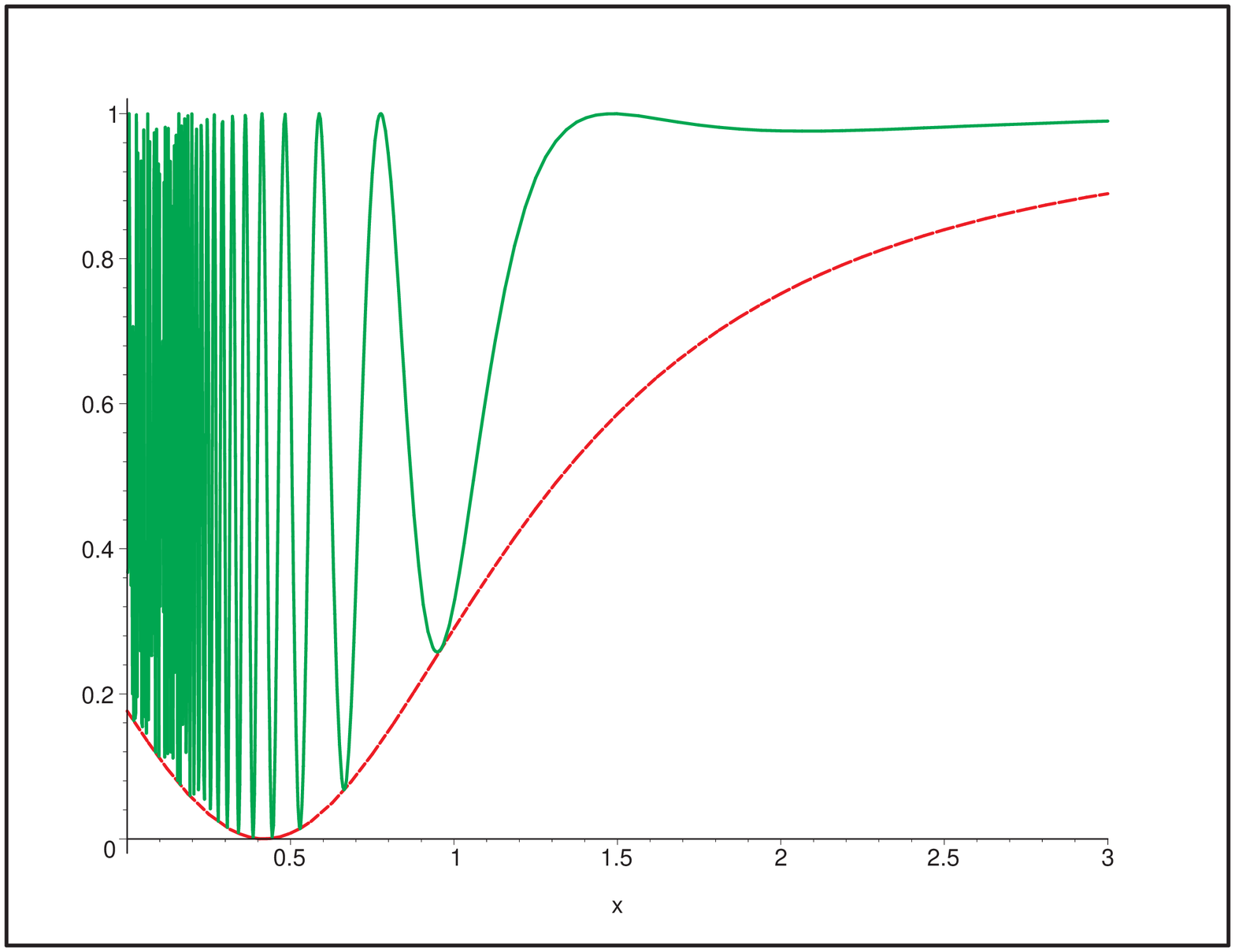}\hfil}
\caption{~~Resonance enhancement of oscillations in matter with constant density.
Shown is a dependence of the ratio of the final and original fluxes, 
$F/F_0$,  on energy ($x \propto E$) for a thin layer, $L = l_0/\pi$ (left panel)   and thick 
layer $L =  10 l_0/ \pi$ (right panel). $l_0$ is the refraction length. 
The vacuum mixing equals $\sin^2 2\theta = 0.824$.  }
\label{resenh}
\end{figure}
In fig.~\ref{resenh} we show  dependence of the ratio $F(E)/F_0(E)$ on energy for 
thin and thick layers. The ratio has an  oscillatory dependence. The oscillatory curve (green) 
is inscribed in to the resonance curve $(1 - \sin^2 2\theta_m)$ (red). The frequency  of the oscillations  
increases with the length $L$. At the resonance energy, the   oscillations proceed with maximal depths.  
Oscillations are enhanced in the resonance range:  
\be
E = E_R \pm \Delta E_R, ~~~ \Delta E_R = \tan 2\theta~ E_R = \sin 2\theta ~E_R^0, 
\ee 
where $E_R^0 = \Delta m^2/2\sqrt{2} G_F n_e$.  
Several comments: for $E \gg E_R$,  matter suppresses the   oscillation depth;  
for small mixing the resonance layer is 
narrow, and the oscillation length in the resonance is large.  
With increase of the vacuum mixing: $E_R \rightarrow 0$ and $\Delta E_R \rightarrow  E_R^0$.

The oscillations in medium with nearly constant density are realized for neutrinos 
crossing the mantle of the Earth. 

\subsection{MSW: adiabatic conversion}

In non-uniform medium,  density changes on the way of neutrinos: 
$n_e = n_e(t)$. Correspondingly, the Hamiltonian of  system depends on time: 
$H = H(t)$. Therefore, 

(i). the mixing angle changes in the course of  propagation: 
$\theta_m = \theta_m (n_e(t))$;

(ii). the (instantaneous) eigenstates of the Hamiltonian, $\nu_{1m}$ and  $\nu_{2m}$,  are no more the 
``eigenstates" of propagation: the transitions $\nu_{1m} \leftrightarrow \nu_{2m}$ occur.

However, if the density changes slowly enough (the adiabaticity  condition) the transitions 
$\nu_{1m} \leftrightarrow \nu_{2m}$ can be neglected. This is the essence of the 
adiabatic condition: $\nu_{1m}$ and $\nu_{2m}$ propagate independently,  as in  vacuum or 
uniform medium. Therefore dynamical features can be summarized in the following way: 

\begin{itemize}
 
\item
The flavors of the eigenstates change according to density change. 
The flavor composition of the eigenstates is determined by $\theta_m(t)$. 

\item
The admixtures of the eigenstates in a propagating neutrino state do not change 
(adiabaticity: no $\nu_{1m} \leftrightarrow \nu_{2m}$ transitions). The admixtures
are given by the  mixing in  production point, $\theta_m^0$. 

\item
The phase difference increases; the  phase velocity is determined by the level splitting  
(which in turn,  changes with density (time)).  

\end{itemize}

Now two degrees of freedom become operative: the relative phase and 
the flavors of neutrino eigenstates. 
The MSW effect is  driven by the  change of  flavors of the  neutrino 
eigenstates in matter with varying density.  
The change of phase produces the  oscillation effect on the top of the adiabatic conversion.\\

Let us comment on the adiabaticity condition. 
If external conditions (density) change slowly,  
the system (mixed neutrinos) has time to adjust this change. 
In general,  the adiabaticity condition can be written as \cite{mess,msad}
\be
\gamma = \left| \frac{\dot{\theta}_m}{H_{2m}  - H_{1m}} \right| \ll 1. 
\label{adiab}
\ee
As follows from the evolution equation for the neutrino eigenstates \cite{ms3,mess}, 
$|\dot{\theta}_m|$ determines the energy of transition 
$\nu_{1m} \leftrightarrow \nu_{2m}$ and  $|H_{2m}  - H_{1m}|$ gives the energy gap between levels. 
The condition (\ref{adiab}) means that 
the transitions $\nu_{1m} \leftrightarrow \nu_{2m}$ can be neglected 
and the eigenstates propagate independently 
(the angle  $\theta_a$  (\ref{exp}) is constant).

The adiabaticity condition is crucial in the resonance layer where 
(i) the level splitting is small and (ii) the mixing angle changes rapidly. 
If the vacuum mixing is small, the adiabaticity is  critical in the resonance point. 
It takes the form \cite{ms1}
\be
\Delta r_R > l_R, 
\label{adiab2}
\ee
where $l_R = l_{\nu}/\sin 2\theta$ is the oscillation length in resonance, 
and $\Delta r_R = n_R/ (dn_e/dr)_R \tan 2\theta$ is the spatial width of resonance layer. 
According to  (\ref{adiab2})  for the adiabatic evolution at least one 
oscillation length should be  obtained in the resonance layer. The adiabaticity condition has been considered outside the 
resonance and in the non-resonance channel in \cite{ssb}. 

In the case of large vacuum mixing the point of maximal adiabaticity violation~\cite{lisi,fried} is shifted 
to densities $n_e(av)$ larger than the resonance density: 
$n_e(av) \rightarrow n_B > n_R$. Here   
$n_B = \Delta m^2 /2\sqrt{2} G_F E$ is the density at  the border of resonance layer for 
maximal mixing. \\

Let us describe  pattern of the adiabatic conversion. 
According to the dynamical  conditions, the admixtures of 
eigenstates are determined by the 
mixing in  neutrino production point. This mixing in turn, 
depends on the density in the initial point,  
$n_e^0$,  as compared to the resonance density. Consequently, a picture of the conversion depends on 
how far from the resonance layer (in the density scale) a neutrino is produced. 

\begin{figure}[htb]
\hbox to \hsize{\hfil\epsfxsize=14cm\epsfbox{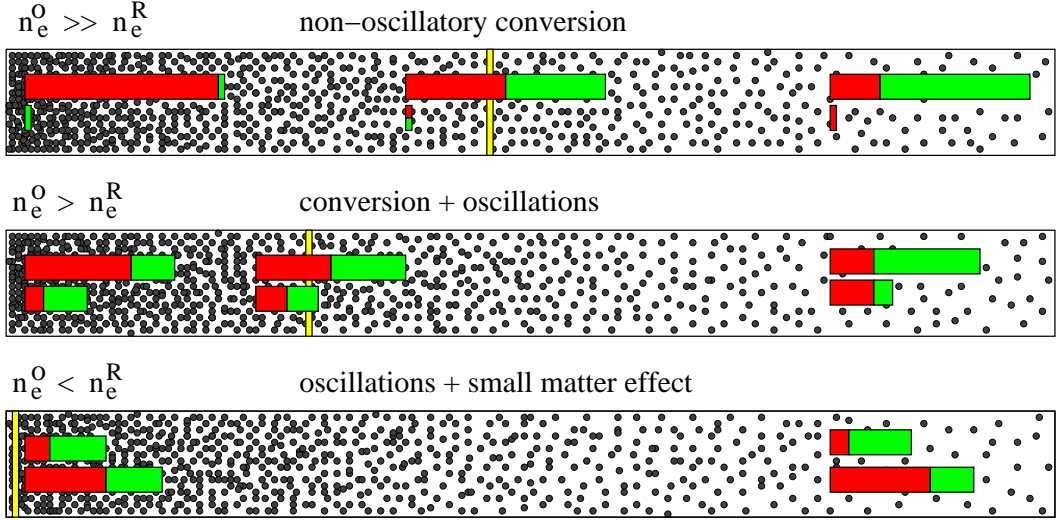}\hfil}
\caption{~~Adiabatic  evolution of  neutrino state for 
three different initial condition ($n_e^0$). 
Shown are the neutrino states  in different moments of propagation in medium 
with varying (decreasing) density. The yellow vertical line indicates 
position of resonance. The initial state is $\nu_e$ in all the cases.  
The sizes of the boxes do not change, whereas the flavors (colors) follow the density change. 
}
\label{faev}
\end{figure}

Three possibilities can be identified.  
They are shown in fig.~\ref{faev}.   
and correspond to large vacuum mixing which is relevant for 
the solar neutrinos. We show propagation of the state 
produced as $\nu_e$ from large density region  to zero density. 
Due to adiabaticity  the sizes of  boxes which correspond 
to the neutrino eigenstates do not change. 

1).  $n_e^0 \gg n_e^R$ - production far above the resonance (the upper panel).  
The initial mixing is strongly suppressed, 
consequently,   the neutrino state, $\nu_e$,  consists mainly of one 
($\nu_{2m}$)  eigenstate, and furthermore, one flavor dominates in a given  
eigenstate.  In the resonance (its position is marked by the yellow line) the mixing is maximal: 
both flavors are present equally. Since the admixture of the second eigenstate is very small, oscillations 
(interference effects) are strongly suppressed. So, here we deal with 
the non-oscillatory flavor transition when the flavor of whole state (which nearly coincides with 
$\nu_{2m}$) follows the density change. At zero density we have $\nu_{2m} = \nu_{2}$,  and therefore 
the probability to find the electron neutrino (survival probability) equals
\be
P = |\langle \nu_e | \nu(t) \rangle|^2 \approx |\langle \nu_e |\nu_{2m}(t) \rangle|^2
= |\langle \nu_e |\nu_{2} \rangle|^2 \approx \sin^2 \theta. 
\label{surv}
\ee
The value of final probability, $\sin^2 \theta$, is the feature of the non-oscillatory transition. 
Deviation from this value indicates a presence of oscillations. 

2).  $n_e^0 > n_e^R$ production above the resonance (middle panel).
The initial mixing is not suppressed.  Although $\nu_{2m}$ is the main component, 
the second eigenstate, $\nu_{1m}$,  
has appreciable  admixture;   
the flavor mixing  in the neutrino eigenstates is significant. 
So, the  interference effect is not suppressed. 
As a result,  here an interplay of the adiabatic conversion and 
occurs. 

3). $n_e^0 < n_e^R$: production below the resonance (lower panel). 
There is no  crossing of the resonance region.  
In this case the matter effect  gives only corrections to the vacuum oscillation picture.

The resonance density is inversely propotional to the 
neutrino energy: 
$n_e^R \propto 1/E$. So, for the same density profile, the condition 
1) is realized for high energies, regime 2) for 
intermediate 
energies and 3) -- for low energies. As we will see all three case are realized for solar neutrinos. 

\subsection{Universality}

The adiabatic transformations show universality:  The averaged probability 
and the depth of oscillations in a given moment 
of propagation are determined by the density in a given point and by 
initial condition (initial density and flavor). They do not depend on 
density distribution between the initial and final points.  
In contrast, the phase of oscillations is an  integral effect of  previous evolution and it depends on 
a density distribution. 

Universal character of the adiabatic conversion can be further generalized in terms of 
variable~\cite{ms3,ms4}  
\be
n = \frac{n_e^R - n_e}{\Delta n_e^R}
\label{y-var}
\ee
which is the distance (in the density scale) from the resonance density  in the units of 
the width of resonance layer. In terms of $n$ the conversion pattern 
depend only on initial value $n_0$. 
\begin{figure}[htb]
\hbox to \hsize{\hfil\epsfxsize=10cm\epsfbox{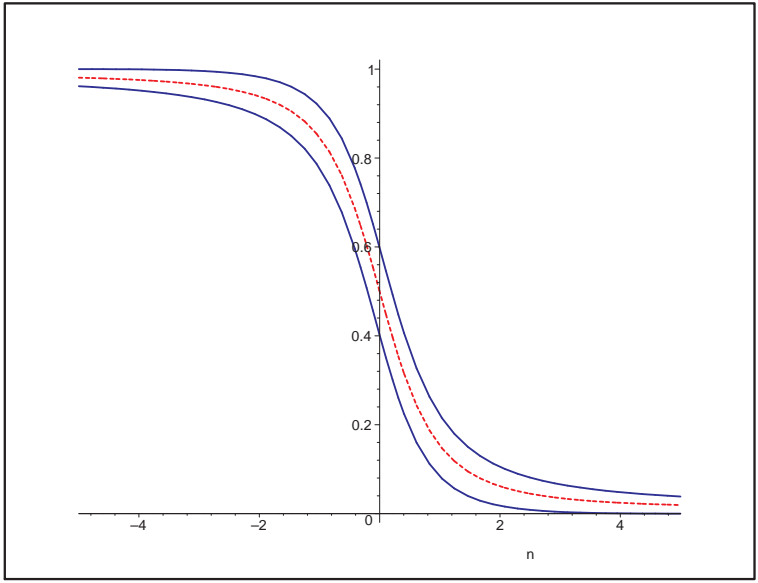}\hfil}
\caption{~~The dependence of the average probability (dashed line) and the depth of oscillations 
($P^{max}$, $P^{min}$ solid lines) on $n$  for  $n_0 = - 5$. 
The resonance layer corresponds to $n = 0$. For $\tan^2 \theta = 0.4$ (large mixing MSW solution) 
the evolution stops at $n_f = 0.47$. 
}
\label{msw}
\end{figure}

In fig.~\ref{msw}   we show  dependences of the average probability and depth of oscillations, that is,  
$\bar{P}$,  $P^{max}$, and $P^{min}$,  on $n$. The probability itself is the oscillatory function which 
is inscribed into the band shown by solid lines. The average probability is shown by the dashed line.  
The curves are determined by initial value 
$n_0$ only, in particular,  there is no explicit dependence on the vacuum mixing angle. 
The resonance is at $n = 0$ and  the resonance layer is given by the interval 
$n = -1 \div  1$. The figure corresponds to $n_0 = - 5$,  {\it   i.e.}, to production  
above the resonance layer; the oscillation depth is relatively small.  
With further decrease of  $n_0$, the oscillation band  becomes narrower 
approaching the line of  non-oscillatory conversion. 
For zero final density we have 
\be
n_f = \frac{1}{\tan 2\theta}. 
\ee
So,  the vacuum mixing enters final condition. 
For the best fit LMA point,  $n_f = 0.45 - 0.50$, and  the evolution should stop at this point. 
The smaller mixing the larger final $n_f$ and the stronger transition.

\subsection{Adiabaticity violation}

In the adiabatic regime the probability of transition between the eigenstates is exponentially 
suppressed  $P_{12} \sim exp{(- \pi/2\gamma)}$ and $\gamma$ is given in (\ref{adiab}) \cite{hax,par}. 
One can consider such a transition as penetration  through a barrier of the height 
$H_{2m} - H_{1m}$  by a  system  with the kinetic energy $d\theta_m /dt$.

If density changes rapidly, so that the condition (\ref{adiab}) is not satisfied, 
the transitions $\nu_{1m} \leftrightarrow \nu_{2m}$ become efficient. 
Therefore  admixtures of the eigenstates in a given propagating state 
change. In our pictorial representation (fig. 5) the sizes of boxes change. 
Now all three degrees of freedom of the system become operative.  

Typically, adiabaticity breaking leads to weakening of the  flavor  
transition.  The non-adiabatic transitions can
be  realized inside  supernovas for the very small 1-3 mixing.

\section{Solar Neutrinos. Large Angle MSW solution}

The first KamLAND result~\cite{kam} has confirmed the large mixing MSW (LMA) solution  of the solar neutrino 
problem. Both the total rate of events and the spectrum distortion are in a very good agreement with predictions 
made on the basis of  LMA \cite{pred}. 

According to the large angle MSW  solution,  inside the Sun the initially produced electron neutrinos 
undergo adiabatic conversion.  Adiabaticity condition is fulfilled  with very high accuracy for 
all relevant energies. Inside the Sun several thousands of oscillation lengths are obtained. 

On the way from the Sun to the Earth the coherence of  
neutrino state is lost and  at the surface of the Earth,  incoherent fluxes of the 
mass states $\nu_1$ and $\nu_2$  arrive. In the matter of the Earth $\nu_1$ and $\nu_2$ 
oscillate producing partial regeneration of the $\nu_e$-flux.

\begin{figure}[htb]
\hbox to \hsize{\hfil\epsfxsize=7cm\epsfbox{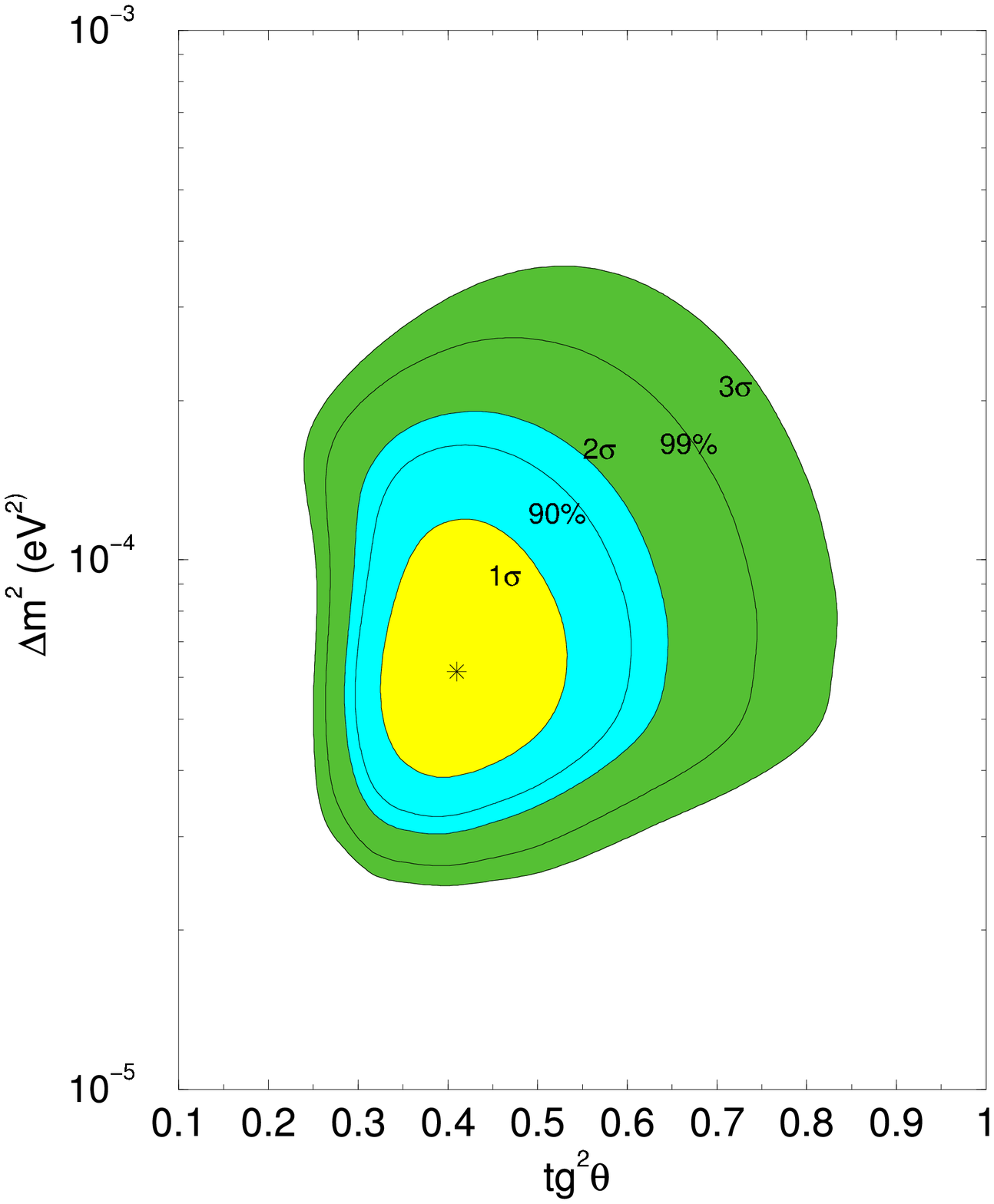}\hfil\epsfxsize=7cm\epsfbox{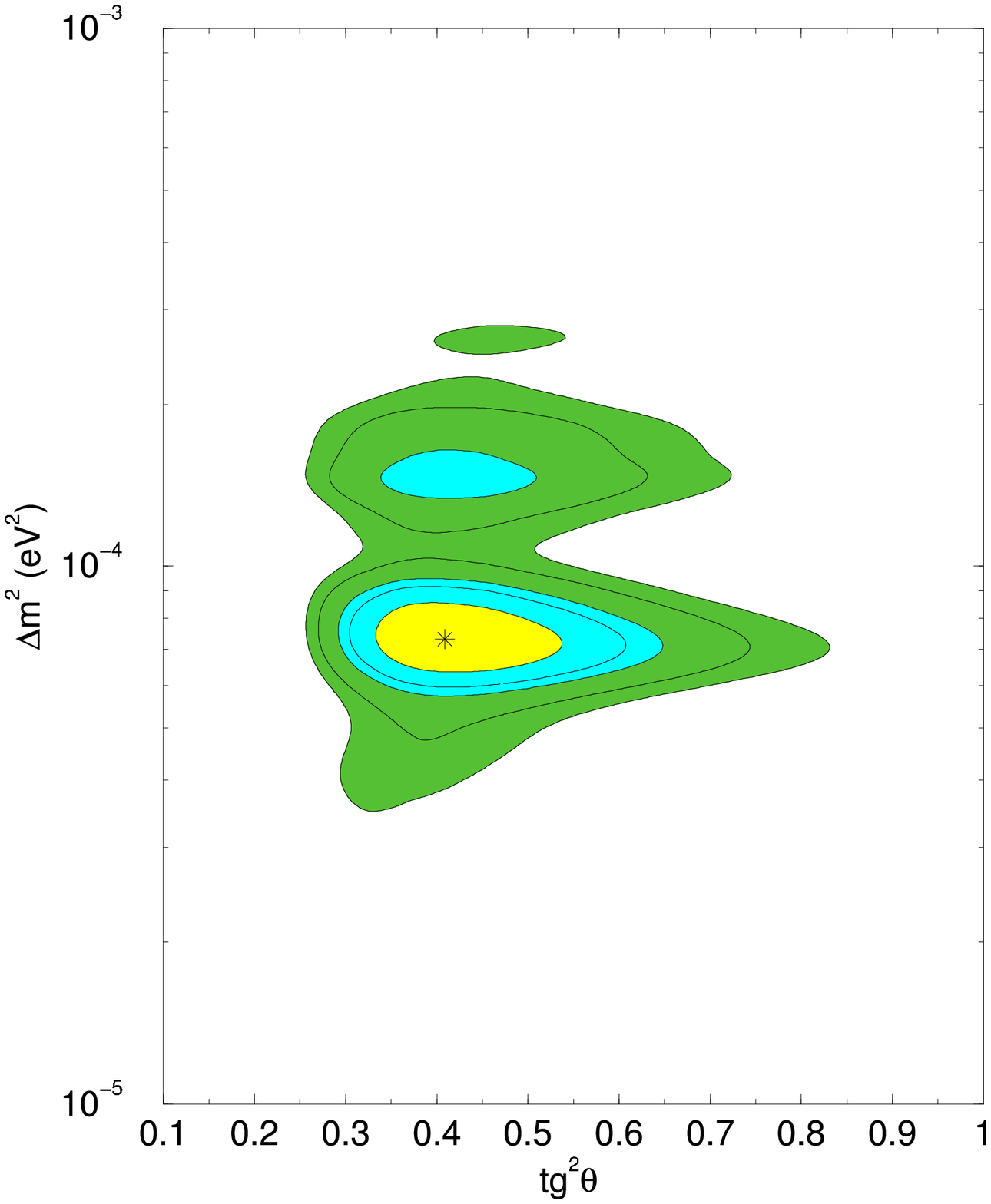}\hfil}
\caption{~~The best fit points and the allowed regions (at different C.L.) 
of the oscillation parameters (at different C.L.)  from the global  fit 
of the solar neutrino data (left), and from the combined analysis of the solar 
neutrino data and KamLAND (right).
}
\label{solution}
\end{figure}

\begin{figure}[htb]
\hbox to \hsize{\hfil\epsfxsize=9cm\epsfbox{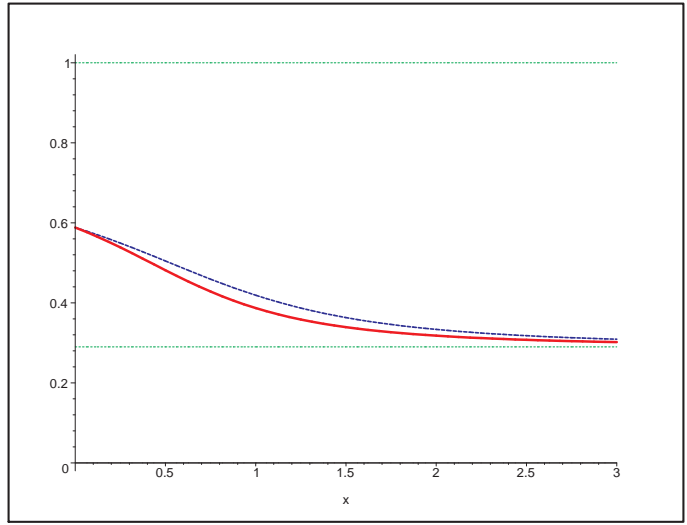}\hfil}
\caption{~~Profile of the effect. Dependence of the survival probability on   
neutrino energy  for the best fit values of parameters and production 
in the center of the Sun (solid line). For $\Delta m^2 = 7\cdot 10^{-5}$ eV$^2$,  $x = 2$ corresponds to 
$E \approx 10$ MeV. The dashed line shows the averaging effect over the production region 
$R =  0.1 R_{sun}$. The Earth matter regeneration effect is not included.} 
\label{profile}
\end{figure}
\begin{figure}[htb]
\hbox to \hsize{\hfil\epsfxsize=7cm\epsfbox{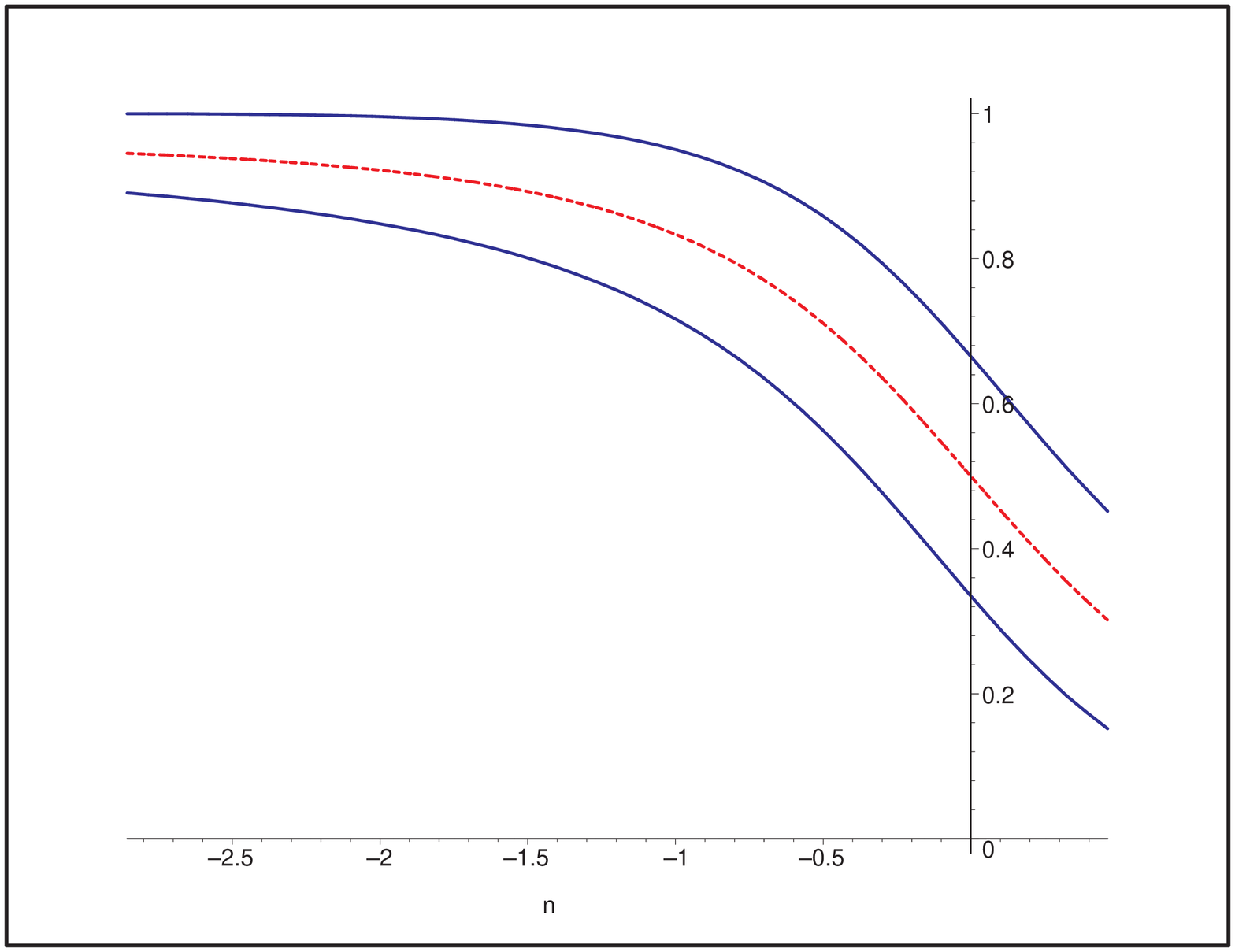}\hfil\epsfxsize=7cm\epsfbox{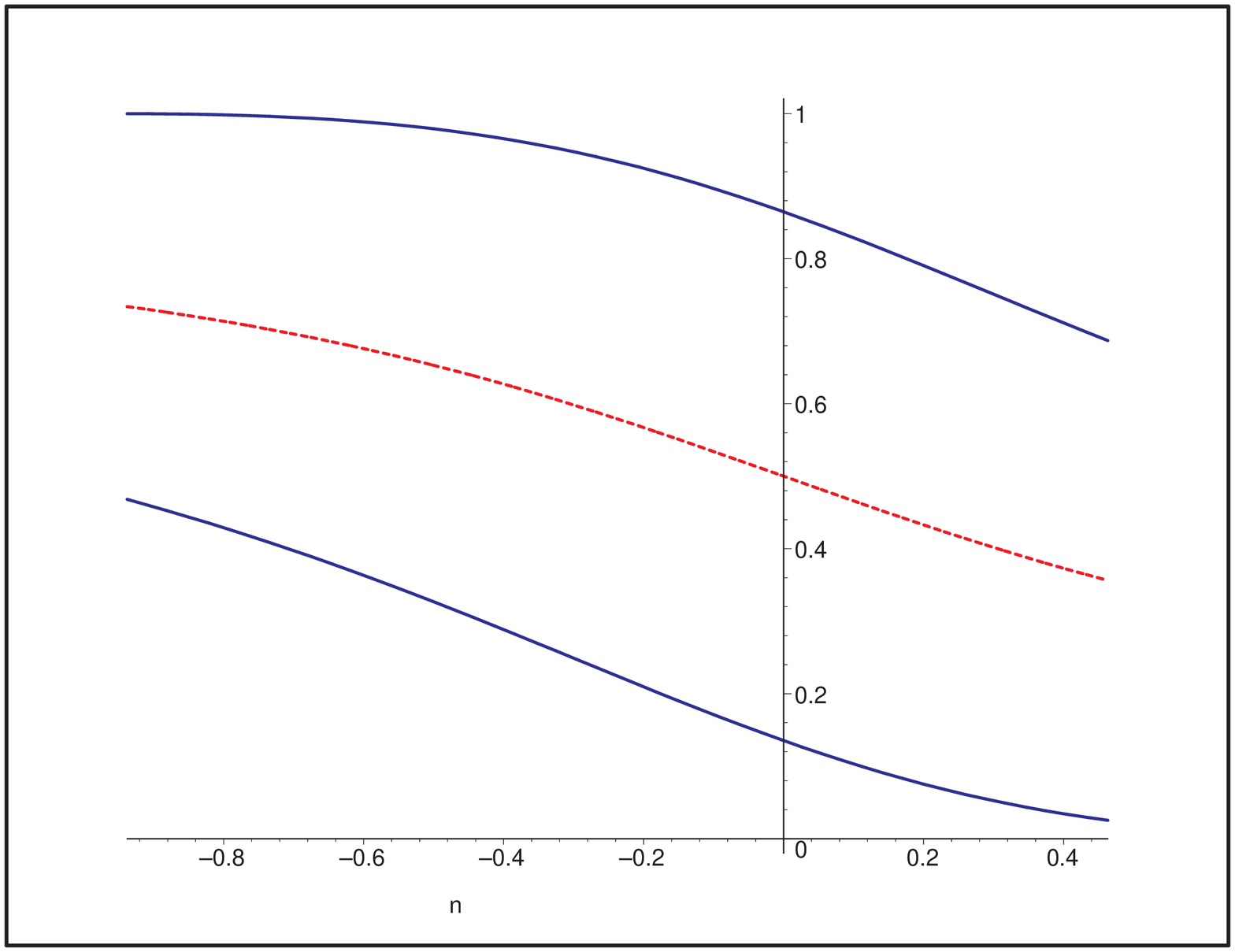}\hfil}
\vskip 0.5cm
\hbox to \hsize{\hfil\epsfxsize=7cm\epsfbox{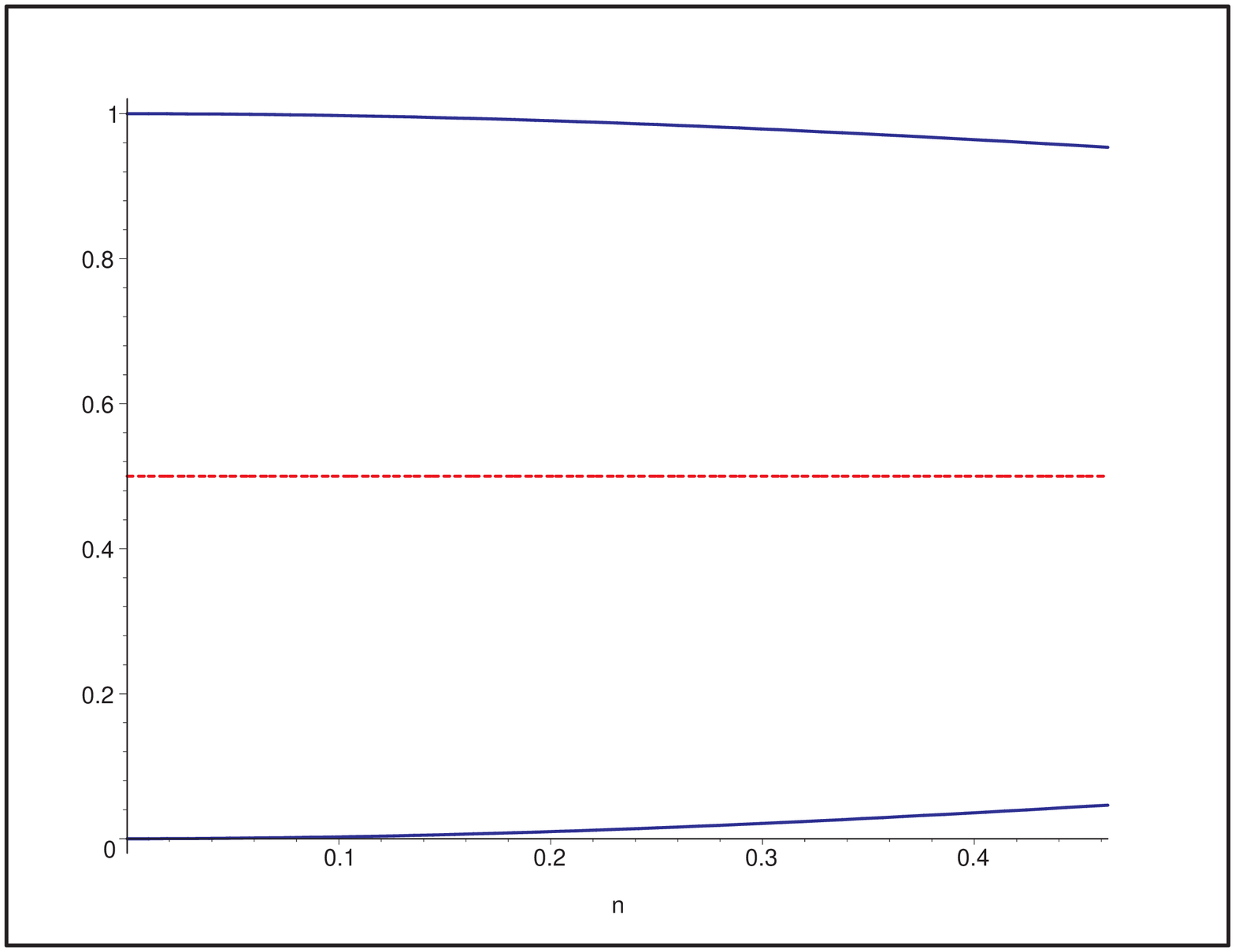}\hfil\epsfxsize=7cm\epsfbox{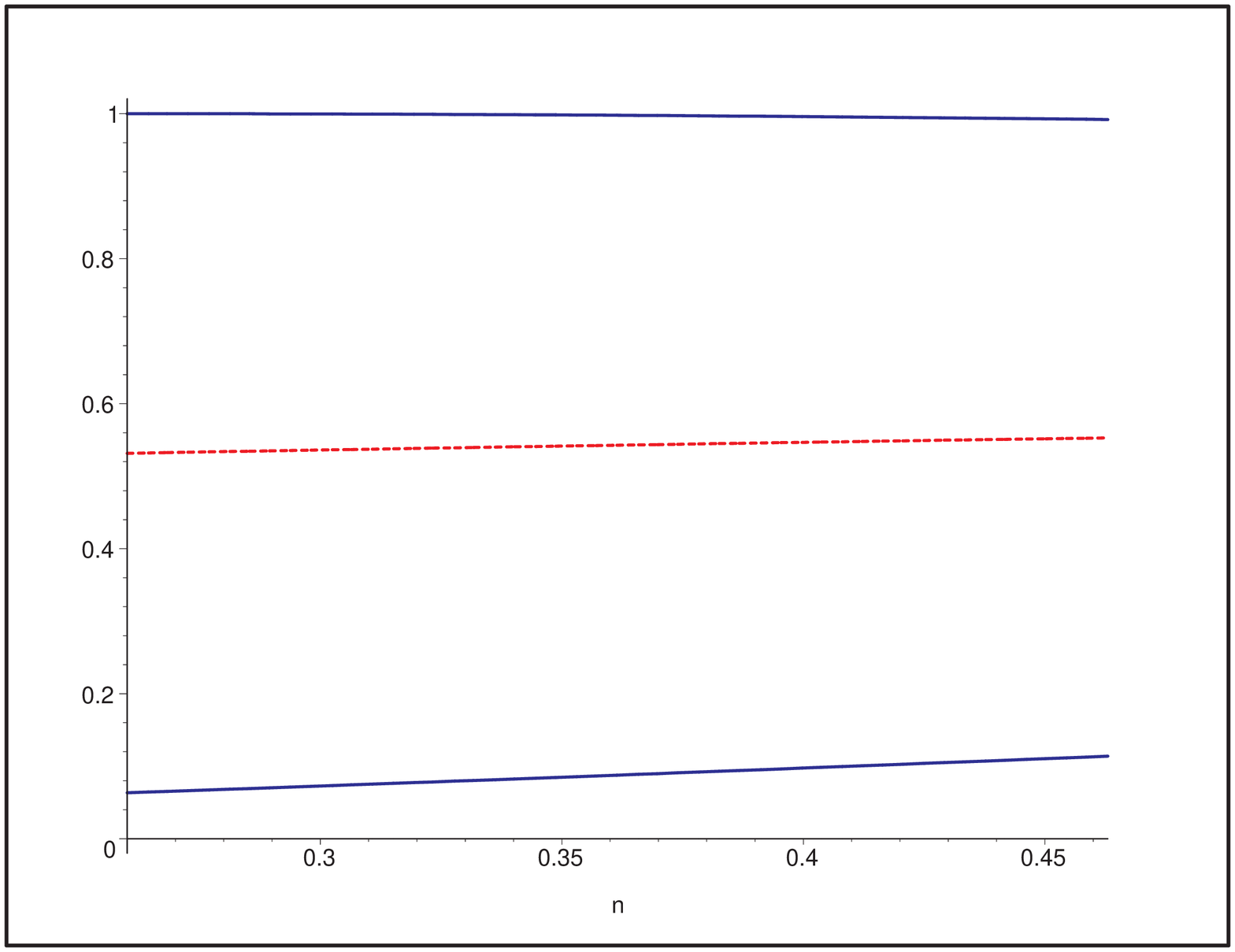}\hfil}
\caption{~~The evolution of the neutrinos with different energies in the Sun. 
Shown are  the dependences of the averaged value of the survival 
probability (dashed lines), as well as maximal and minimal values of the probability (solid lines) 
on  $n$. Neutrino  is produced in the center of  the Sun. The density decreases from the left to the 
right.  The resonance cis at $n = 0$. The probability is the oscillatory 
curve which is inscribed in the band between $P^{max}$ and $P^{min}$. 
Upper left panel: $E = 14$ MeV,  upper right panel: $E = 6$ MeV, lower left panel $E = 2$ MeV,  
lower right panel $E = 0.86$ MeV.}
\label{pattern}
\end{figure}

In fig.~\ref{solution} from \cite{comb} we show  the best fit point and the allowed regions of 
oscillation parameters 
from  (a)  analysis of  the solar neutrino data   and (b) 
combined analysis of the solar and KamLAND results (in assumption of the CPT invariance). 
The best fit point is at 
\be
\Delta m^2 \sim 7 \cdot 10^{-5} {\rm eV}^2, ~~~~ \tan^2 \theta \sim 0.4. 
\label{param}
\ee 
For these parameters,  
the energy ``profile of the effect" - 
the dependence of the survival probability 
on the neutrino energy is shown in fig.~\ref{profile}.  In fig.~\ref{pattern} we present  conversion patterns 
for different neutrino energies. 


There are  three energy ranges with different features of transition: 

1. In the high energy part of spectrum, $E > 10$ MeV ($x > 2$), the  adiabatic conversion with small 
oscillation effect occurs. The spatial evolution (in $n$-scale) is shown in the upper left panel of  fig.~\ref{pattern}. 
At the exit, the resulting averaged probability is slightly larger than $\sin^2 \theta$ expected 
from  the non-oscillatory transition.  
Pictorial representation of the conversion is shown in fig. \ref{pattern}.  
With decrease of energy the initial density approaches the resonance density, 
and the depths of oscillations increases. 

2. Intermediate energy range $E \sim (2 - 10)$ MeV ($x = 0.3 - 2$ ) the oscillation effect is significant. 
The interplay of the oscillations and conversion takes place (fig.~\ref{pattern}). 

For $E \sim 2$ MeV  neutrinos are produced in resonance (bottom left panel). Initial depth of oscillations is  
maximal and $\bar{P} = 0.5$.

3. At low energies: $E <  2$ MeV ($x < 0.3$),  the  vacuum oscillations  with small matter corrections 
occur.  

Notice that without matter effect for all energies one would  get the pattern of evolution 
close to that in the bottom-right panel. 

Basically this is what was called the ``adiabatic solution" in the early days: 
the boron neutrino spectrum 
is ``sitting" on the adiabatic edge of the suppression pit. An absence of observable spectrum distortion 
allows now only the large mixing part of the adiabatic solution.

As a specific example, let us consider  neutrinos with $E = 10$  MeV produced in the center of the 
Sun. For these neutrinos the resonance density equals $n_R Y_e = 20$ g/cc, where $Y_e$  is the 
number of electrons per nucleon. The resonance layer is in the central parts of the Sun: 
$R_R = 0.24 R_{sun}$. In the production point: 
$\sin^2 \theta_m = 0.94$ and $\cos^2 \theta_m = 0.06$, so indeed, $\nu_{2m}$ dominates.  
At the surface of the Sun the  state  $\nu_{2m}$  appears as $\nu_{2}$ and then arrives at the Earth 
loosing the coherence with $\nu_{1}$. 
Entering the Earth the state $\nu_{2}$ splits in two matter eigenstates: 
\be
\nu_{2} \rightarrow \cos\theta_m'  \nu_{2m} + \sin \theta_m'\nu_{1m}.
\ee  
It oscillates regenerating the  $\nu_e$-flux. With the Earth matter effect taken into account,  
the survival probability  at high energies can be written as 
\be
P \approx \sin^2 \theta + f_{reg},  
\ee 
where the regeneration factor  equals
\be
f_{reg} = 
0.5 \sin^2 2\theta ~\frac{l_{\nu}}{l_0}~.  
\label{freg}
\ee
Notice that the oscillations of $\nu_{2}$  are  pure matter effect and for the  presently favored 
value of $\Delta m^2$ this effect is small. According to (\ref{freg}), 
$f_{reg} \propto 1/\Delta m^2$ and the expected day-night asymmetry  of the charged current signal 
equals
\be
A_{DN} = f_{reg}/P  \sim (3 - 5)\%~. 
\label{freg1}
\ee

\section{Beyond LMA}

Is the solar neutrino problem solved? 
In assumption of the CPT,  the large angle  MSW effect is indeed  the dominant mechanism 
of the solar neutrino conversion, and all other possible mechanisms could 
give the sub-dominant effects only.

What is the next? First of all, even accepting  the LMA solution  one needs better determination of the 
oscillation parameters. In particular,  further improvements of the upper bounds on  
$\Delta m^2$ and $\tan^2 \theta$ (deviation from maximal mixing) are  of great importance.  
These improvements have  implications for both phenomenology (future long baseline experiments,  
double beta decay searches, {\it etc.}) and theory.  
The improvements are also needed for better understanding of  physics of the solar neutrino conversion. 
In sect. 6 we have described  the  picture which corresponds to the oscillation parameter near the best fit point 
(\ref{param}). Physics, in particular relative importance of the vacuum oscillations and the matter effect, 
changes with parameters within the presently allowed region (fig. \ref{solution}).

Is large mixing MSW sufficient to describe the data? If there are  observations 
which may indicate  some deviations from LMA? 
According to recent analysis, LMA describes all the data very well: pulls 
of predictions from results of measurements  are below $1\sigma$ 
for all but one experiment~\cite{comb}). 
High Ar-production rate,  $Q_{Ar} \sim 3$ SNU, is a generic prediction of LMA.  
The predicted rate is about $2\sigma$ above the Homestake result. 
This difference can be  statistical fluctuation or some systematics which may  be  
related to the claimed time variations of the Homestake signal. 

Another generic prediction  is  the  ``turn up of  spectrum"  (spectrum distortion) at low energies.  
According to LMA the survival probability should increase with decrease of energy 
(fig.~\ref{profile}):  for the best fit point the turn up can be as large as 10 - 15\% 
between 8 and 5 MeV~\cite{comb}. 
Neither SuperKamiokande nor SNO show any turn up although the present sensitivity is not enough to 
make any statistically significant statement. 

Are these observations  related? Do they indicate some new physics 
at low  energies? It happens that both the lower $Ar$-production 
rate and absence of (or weaker)  turn up of the 
spectrum can be explained by the  effect of new (sterile) neutrino~\cite{pedr}. 

Suppose that on the top of usual  pair of states with the LMA parameters  
(\ref{param}) new light neutrino state, $\nu_s$,  exists which 

- mixes weakly with the lightest state $\nu_1$: $\sin^2 2\theta_{01} \sim (10^{-4} - 10^{-3})$,

- has the mass difference with $\nu_1$: $\Delta m^2_{01} = (2 - 10) \cdot 10^{-6}$ eV$^2$. 
If $\nu_1$ is very light, the mass of $\nu_0$  equals  $(2- 3)\cdot 10^{-3}$ eV.

It can be shown, that the presence of such a  neutrino does not change the survival probability 
in the non-oscillatory and vacuum ranges but do change it in the transition region. 
In general, it leads to  appearance of a dip in the adiabatic edge: 
at $E = (0.5 - 2)$ MeV and flattening of spectrum distortion at higher energies. 
The dip produces  suppression of the $Be$-neutrino flux as well as other  fluxes 
at the intermediate energies, and consequently, 
suppression of the $Ar$-production rate. It also diminishes or eliminates completely (depending on the 
angle and $\Delta m^2_{01}$) the turn up of spectrum.  
This scenario predicts low rate in BOREXINO: it can be as low as $\sin^4 \theta \sim 0.1$ of the SSM rate. 
The scenario implies also  smaller 1-2 mixing, $\tan^2 \theta$, (to compensate decrease of the $Ge$-production rate) 
and larger boron neutrino flux (to reproduce the high energy data). 
For $\Delta m^2 > (1 - 2)\cdot 10^{-5}$ eV$^2$ the turn up can be changed without diminishing 
the $Be$- neutrino flux. 

Smallness of  mixing of the  sterile neutrino allows to avoid the nucleosynthesis bound: 
such a neutrino does  not equilibrate in the Early Universe.

\section{Summary}

1. We have described here two matter effects: 

The resonance enhancement of oscillations in matter with constant density. 

The adiabatic (quasi-adiabatic) conversion in medium with varying density (MSW).\\

2. Adiabatic (quasi-adiabatic) conversion is related to the  change of the mixing in matter 
on the way of neutrino, or equivalently, to the change of flavors of the neutrino 
eigenstates. 
In contrast,  oscillations are related to  change of the relative phase of the eigenstates.\\

3. The large mixing MSW effect provides the solution of the solar neutrino problem. 
The solar neutrino data allow to determine  the oscillation parameters $\Delta m^2_{12}$ and 
$\theta_{12}$ and therefore to make next important step in reconstruction of 
the neutrino mass and flavor spectrum.

Now we can say how the mechanism of  conversion of the solar neutrinos works. 
A picture of the conversion depends on neutrino energy.  It has a character of 

- nearly non-oscillatory transition for $E > 10$ MeV, 

- interplay of the adiabatic conversion and oscillations for $E = 2 - 10$ MeV,  

- oscillations with small matter corrections for $E < 2$ MeV. \\

4. The large angle MSW effect is the dominant mechanism of the  solar neutrino conversion. 
Although more precise determination  of parameters is needed to identify completely 
the physical picture of the effect.  All other suggested mechanisms can produce  sub-leading effects.
With the available  data we know 
rather well what happens with high  energy neutrinos ($ E > 5$ MeV). Still some physics beyond  LMA  may 
show up  in the  low energy  part of spectrum. The low $ Ar$ -production rate and absence of the turn up 
of the spectrum  distortion in the range $E < 8$ MeV 
can be  due to  an additional effect of the light sterile neutrino with very small mixing. 
BOREXINO~\cite{bor} and KamLAND can test this possibility in future.

\section{Acknowledgments}

I would like to thank Milla Baldo-Ceolin for invitation to give this talk and for hospitality  
during my stay in Rome and Venice.

\end{document}